\documentclass[preprint,3p,11pt]{elsarticle}
\usepackage{graphicx}
\usepackage{epsfig}
\usepackage{amsfonts}
\usepackage{amsmath,amssymb}
\journal{Physics Letters B}

\begin{document}
\begin{frontmatter}
\title{Disentangling phase transitions and critical points in the
  proton-neutron interacting boson model by  catastrophe theory}

\author[1]{J.E. Garc\'{\i}a-Ramos\corref{cor1}}
\ead{enrique.ramos@dfaie.uhu.es}

\author[2]{J.M. Arias}
\ead{ariasc@us.es}

\author[3]{J. Dukelsky}
\ead{dukelsky@iem.cfmac.csic.es}

\address[1]{Departamento de F\'{\i}sica Aplicada, Universidad de Huelva,
21071 Huelva, Spain, \\ and Unidad Asociada de la Universidad de Huelva
al IEM (CSIC), Madrid, Spain.}

\address[2]{Departamento de F\'{\i}sica At\'omica, Molecular y Nuclear,
Universidad de Sevilla, Apdo~1065,
41080 Sevilla, Spain,\\and Unidad Asociada de la Universidad de Sevilla
al IEM (CSIC), Madrid, Spain.}

\address[3]{Instituto de Estructura de la Materia, CSIC, Serrano 123,
28006 Madrid, Spain.}

\cortext[cor1]{Corresponding author: Tel.~+34959219791,
   Fax:~+34959219777}

\begin{abstract}
We introduce the basic concepts of catastrophe theory  needed to derive analytically the phase diagram of the proton-neutron
interacting boson model (IBM-2). Previous studies \cite{Ar04,CI04,Caprio05} were based on numerical solutions. We here explain the whole IBM-2 phase diagram including the precise order of the phase transitions in terms of the {\it cusp catastrophe}.\\
\end{abstract}
%\maketitle


\end{frontmatter}

\section{Introduction}
\label{sec-intro}
In the last twenty years quantum phase transitions (QPT's) (phase transitions
that happen at zero temperature as a function of a control parameter)
has been a subject of great interest in different areas of quantum many-body systems. In particular, in Nuclear Physics the study of shape phase
transitions is a topic of current interest both theoretical and experimentally
\cite{Cast07,Cejn09,Cejn10}. Moreover, in Molecular Physics
\cite{Iach08,Pere11}, Quantum Optics \cite{Emar,Amic08} and Solid State Physics
\cite{Sach11} the interest on QPT's has grown enormously in recent years.

Strictly speaking, QPT's take place for large
systems in the thermodynamic limit as a discontinuity or singularity in some derivative of the ground state energy. However, finite systems like the atomic nuclei could show the
precursors of a phase transition  when structural changes in the ground state are observed as a function of the neutron or proton numbers (N,Z) \cite{IZ04}. These transitional nuclei are characterized by specific patterns in the low lying spectrum which could be associated with a critical system. Recently, in
the context of the Bohr Hamiltonian \cite{BMII},  F.\ Iachello has
introduced the concept of critical point symmetry \cite{E5,X5,Y5} that
can be applied when the quantum system undergoes transitions between two
different shape-phases.

A convenient model to study QPT's in nuclei is the interacting boson model
(IBM) \cite{IBM}. It is a symmetry dictated model whose building
blocks are ideal bosons representing nucleon pairs, with angular
momentum zero, $s$ bosons, and two, $d$ bosons.
In its simplest version, IBM-1, the
dynamical algebra of the model is U(6) and the common symmetry algebra
O(3). IBM-1 presents three dynamical symmetries (SU(5), O(6), and SU(3))
corresponding to well defined nuclear shapes (spherical,
$\gamma$-unstable, and axially deformed,
respectively). The presence of different dynamical
symmetries in the model is a key ingredient to study QPT's, since they
appear when two different symmetries are mixed in the Hamiltonian through a
control parameter:
$
%\begin{equation}
H(\xi)=\xi\cdot H(\mbox{symmetry}_1)+ (1-\xi)\cdot H(\mbox{symmetry}_2).
$
%\end{equation}
For a particular value of the control parameter, $\xi_c$, the system
undergoes a structural QPT from symmetry 1 to symmetry 2.

The geometry and shape phase transitions of the IBM-1
were studied long ago  \cite{DSI80,FGD,Alex} and also more recently
the many facets of QPT's in IBM have been analyzed \cite{Cejn09}.
%\cite{Lope96,Jo1,Ar3}.
A similar kind of study could be extended to the more realistic version of the IBM that
includes the proton-neutron degree of freedom, known as
IBM-2 \cite{IBM-2}. Most of the studies \cite{Ar04,CI04,Caprio05} rely
on the numerical diagonalization of the IBM-2 Hamiltonian which is
restricted to systems with, as maximum, $10$ proton and $10$ neutron bosons, or in the numerical treatments of semiclassical approximations. These numerical studies, that provided a very accurate description of the phase diagrams,
could not state in an unambiguous way the classification of phase transition orders.
In this letter we present an analytic study of the IBM-2 phase
diagram that is able to determine unambiguously the order of the phase transitions
present in the two-fluid IBM-2 model making use of the catastrophe
theory (CT) \cite{Thom75,Post78,Gilm81}. CT
allows to analyze energy surfaces
depending on various parameters, {\it i.e.}, families of potentials,
that contain several shape
variables, providing information on the nature and numbers of
minima (also maxima, if they exist), i.e., stability, depth, etc.
This analysis leads to a partition of the parameter space into
different regions  where the energy surface has different qualitative
properties (number and nature of maxima and minima) and, in
particular, it serves to study and classify the existing QPT's. Similar studies were carried out for IBM-1 \cite{Lope96}, for IBM
including configuration mixing \cite{Hell07,Hell09}, for
three-components thermodynamic systems \cite{Gait98,Gait99}, and for
elementary chemical reactions \cite{Marg00}.

\section{IBM-2 Hamiltonian}
\begin{figure}
\begin{center}
\includegraphics[width=12cm]{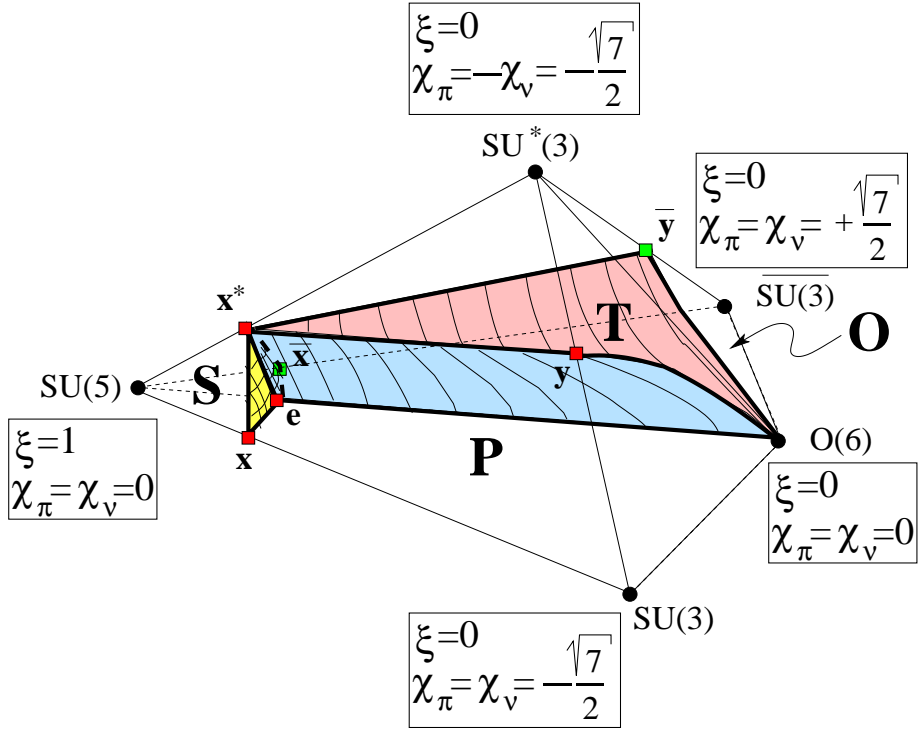}
\end{center}
%\vspace{6cm}
%\vspace{0.7cm}
\caption{Schematic phase diagram for IBM-2 \cite{Ar04,CI04},
  including the prolate and the oblate sides.}
\label{FIG5}
\end{figure}
In this work we use a simplified
form of the IBM-2 Hamiltonian, known as the
Consistent-Q Hamiltonian \cite{Ca}, which retains all the main
ingredients of the full Hamiltonian
\begin{equation}
H= \xi \left(n_{d_\pi}+n_{d_\nu} \right)-\frac{1-\xi}{N}
Q^{(\chi_\pi,\chi_\nu ) }\cdot Q^{(\chi_\pi,\chi_\nu )} ~,
\label{HQ}
\end{equation}
where $n_{d_{\rho}}=\sum_{\mu }d_{\rho \mu }^{\dagger }d_{\rho \mu }$,
$Q^{(\chi_\pi,\chi_\nu )}_{\mu}=\left(Q_\pi^{\chi_\pi}+
Q_\nu^{\chi_\nu}\right)_{\mu}$ with
$Q^{\chi_{\rho}}_{\mu
}=\left[ d^{\dagger }_\rho \widetilde s_\rho +s^{\dagger }_\rho \widetilde d_\rho\right]
_{\mu }^{2} +\chi_\rho \left[ d^{\dagger }_\rho \widetilde{d_\rho}\right] _{\mu
}^{2}$ and $\rho = \pi, \nu$. $N=N_\pi + N_\nu$ is the total number of bosons representing the number of valence nucleon pairs.

We study the phase diagram of the IBM-2 in the semi-classical or mean
field formalism. In this approach, which is exact in the thermodynamic
limit, the ground state wavefunction is a product of a proton condensate times a neutron condensate \cite{GK80, DDS82}, $|g\rangle=|N_\pi, N_\nu, \beta_\pi, \gamma_\pi ,
\beta_\nu, \gamma_\nu \rangle$
\begin{equation}
|g \rangle
= \frac{1}{\sqrt{N_\pi! N_\nu!}} ~\left(\Gamma^\dag_\pi\right)^{N_\pi}
\left(\Gamma^\dag_\nu\right)^{N_\nu} |0 \rangle ~,
\label{STIN}
\end{equation}
where $|0 \rangle$ is the boson vacuum and $\Gamma^\dag_\rho$ is the
creation operator for a coherent $\pi$ or $\nu$ boson, defined as,
\begin{equation}
\Gamma^\dagger_\rho=\frac{1}{\sqrt{1+ \beta_\rho^2}}\left[s_\rho^\dagger + \beta_\rho \cos
    \gamma_\rho d^\dagger_{\rho 0} +  {1 \over \sqrt{2}} ~
\beta_\rho \sin \gamma_\rho (d^\dagger_{\rho 2} + d^\dagger_{\rho -2})
\right]  ~.
\end{equation}

The equilibrium values of the structure parameters
($\beta_\pi, \gamma_\pi , \beta_\nu, \gamma_\nu$) and the energy of
the system for given values of the control parameters ($\xi$, $\chi_\pi$, $\chi_\nu$) in the
Hamiltonian (\ref{HQ}) can be obtained by minimizing
its expectation value in the intrinsic
state (\ref{STIN}): $\delta \langle g |H|g \rangle = 0$.
In the limit $N_\pi,N_\nu \rightarrow \infty $ the energy per boson can be obtained in a straightforward way as

%\begin{eqnarray}
%&&E(\beta_\pi, \gamma_\pi , \beta_\nu, \gamma_\nu;\chi_\pi, \chi_\nu, x)= x
%\left[\frac{\beta_\pi^2}{1+\beta_\pi^2} +
%\frac{\beta_\nu^2}{1+\beta_\nu^2} \right] \nonumber \\
%&-& \frac{1-x}{4} \left[Q_0^2(\beta_\pi,\gamma_\pi,\chi_\pi) \right. \nonumber
%\\
%&+&
%Q_0^2(\beta_\nu,\gamma_\nu,\chi_\nu) + 2
%Q_2^2(\beta_\pi,\gamma_\pi,\chi_\pi) + 2
%Q_2^2(\beta_\nu,\gamma_\nu,\chi_\nu) \nonumber \\
%&+& 2
%Q_0(\beta_\pi,\gamma_\pi,\chi_\pi) Q_0(\beta_\nu,\gamma_\nu,\chi_\nu)
%\nonumber \\
%&+& \left. 4  Q_2(\beta_\pi,\gamma_\pi,\chi_\pi)
%Q_2(\beta_\nu,\gamma_\nu,\chi_\nu)
%\right]
%\end{eqnarray}
\begin{equation}
E(\beta_\pi, \gamma_\pi , \beta_\nu, \gamma_\nu;\chi_\pi, \chi_\nu,
\xi)= \frac{\xi}{2}
\sum_{\rho=\pi,\nu} \frac{\beta_\rho^2}{1+\beta_\rho^2}
-
\frac{1-\xi}{4} \sum_{\mu=0,\pm 2} \left[\sum_{\rho=\pi,\nu}
  \left(Q^{\chi_\rho}_\mu(\rho)\right)^2 +  2 ~
  \left(Q^{\chi_\pi}_\mu(\pi)\right)
  \left(Q^{\chi_\nu}_{-\mu}(\nu)\right) \right]
\label{Ener}
\end{equation}
with
\begin{eqnarray}
Q^{\chi_\rho}_0(\rho)&=&\frac{1}{1+\beta_\rho^2} {\left[ 2 \beta_\rho
  \cos \gamma_\rho   - \sqrt{\frac{2}{7}}
\beta_\rho^2 \chi_\rho \cos (2\gamma_\rho) \right]} \\
Q^{\chi_\rho}_2(\rho) = Q^{\chi_\rho}_{-2}(\rho)
&=&\frac{1}{1+\beta_\rho^2}\left[
\sqrt{2} \beta_\rho \sin \gamma_\rho + \sqrt{\frac{1}{7}}
\beta_\rho^2 \chi_\rho \sin (2\gamma_\rho) \right].
\end{eqnarray}
Notice that Hamiltonian (\ref{HQ}) can only lead to energy surfaces where
proton and neutron ellipsoids are axially symmetric with symmetry
axis either parallel or perpendicular, or to triaxial shapes in both
ellipsoids (see \cite{Gino92} for details). Therefore, it is not required to
include explicitly the Euler angles.

In order to analyze the energy surface it is convenient to introduce new
variables,
\begin{equation}
\chi=\frac{\chi_\pi+\chi_\nu}{2},\qquad \chi'=\frac{\chi_\pi-\chi_\nu}{2}.
\end{equation}
In Fig.~\ref{FIG5} we depict the phase diagram of the model, which has
been obtained numerically \cite{Ar04,CI04,Caprio05}. The different phases correspond to the spherical region $S$,
 the prolate axially deformed region $P$, the oblate axially deformed
 region $O$,  and
the region with triaxial shapes $T$. It is interesting to note that the IBM-1 phase diagram is
recovered for $\chi'=0$.  The phase diagram could also be extended to
positive values of $\chi'$, by reflection in the horizontal plane.
%We have included in Fig.~\ref{FIG5} the region of oblate
%axially deformed shapes (labeled with $O$).

The orders of the different phase transitions were inferred from the numerical calculations
in \cite{Ar04,CI04,Caprio05}. All references coincide in the following classification:
\begin{itemize}
\item the $x-x^*-e-x$ surface:
(equally the surface $\overline{x}-x^*-e-\overline{x}$ in the extended diagram to include oblate
shapes) is first order, except for the $x^*-e$ line, which was
proposed to be second order.
\item the $x^*-e-O(6)-y-x^*$ surface: (equally the surface
  $x^*-e-O(6)-\overline{y}-x^*$ in the extended diagram to include
  oblate shapes) is second order.
\item the line $e-x^*$: is second order.
\item the line $e-O(6)$: is first order.
\end{itemize}

In what follows we will introduce the basic ingredients of CT to
determine in an unambiguous way the properties of the QPTs along the
critical lines/surfaces in the IBM-2 phase diagram.
%In Fig. \ref{fig-path}, the value of the ground
%state energy and the variation of the
%order parameters is depicted for a particular trajectory to show
%how difficult is to find out exactly where a QPT appears and also the
%order of such a transition.
% \begin{center}
% \begin{figure}
% \includegraphics[height=6cm,width=6cm]{c-int-new3.eps}
% %\vspace{6cm}
% %\vspace{0.7cm}
% \caption{Transition from U(5) to a triaxial shape: $\chi_\pi=-1.2$,
%   $\chi_\nu=0.5$ and the control parameter $x$ varies from $0$ to $1$.}
% \label{fig-path}
% \end{figure}
% \end{center}

\section{Catastrophe theory program}
\label{sec-CTP}

Once the IBM-2 phase diagram is known through a numerical
study \cite{Ar04,CI04,Caprio05}, for the conclusive determination
of the order of the phase transitions associated to the obtained critical
lines/surfaces it is required an analytic study. CT is specially suited to carry out such analysis.
%We have seen in section \ref{sec-phase} that a numerical
%study is ambiguous in determining the order of the phase
%transitions, however, catastrophe theory (CT)
%\cite{Zeem77,Gilm81,Arno92} provides an
%ideal framework to determine the structure of the phase diagram, and,
%that is most important, to obtain the order of a phase transition
%\cite{Lope96}.

The basic goal of CT is the study of a potential or, more generically, a family of potentials. As we will see, these potentials
comprise three types of points. To begin with, let us assume a system described by a real family of
smooth potentials:
\begin{equation}
V(\vec{x}, \vec{\lambda})\in\Re
\end{equation}
where $\vec{x} \in\Re^n$ stand for the state (order) variables  and
$\vec{\lambda}\in\Re^r$ are the control parameters.
The majority of points $\vec{x}$ have a nonzero gradient and are
called regular points. The points where the gradient
vanishes are called stationary or critical points and they can be classified in
two groups: i) the points where the determinant of the Hessian matrix  is
different from zero, called isolated, non-degenerated or
Morse points, and ii) the points where the determinant of the
Hessian matrix is zero, called non-isolated, degenerated or non-Morse
points. In summary, points of a family of smooth potentials can be
classified according to their gradient and Hessian matrix ${\cal H}$ as:
\begin{itemize}
\item Regular points: $\nabla V\neq 0$.
\item Morse points (isolated critical points):
$\nabla V= 0$ and $|{\cal H}|\neq 0$.
\item Non-Morse points (degenerated critical points):
$\nabla V= 0$ and $|{\cal H}|= 0$.
\end{itemize}

Morse theorem \cite{Post78,Gilm81}
guarantees that around a Morse point, a smooth potential is
equivalent to a quadratic form, thanks to a smooth non-linear change
of variables.
%a quadratic
%form exits and is equivalent to a family of potentials for a smooth
%change of variables around a Morse point.
Therefore, the stability of the
potential under small perturbation in the parameters is guaranteed in
Morse points.
At non-Morse points the potential cannot be written as a quadratic
form because the Hessian matrix has at least one zero eigenvalue.
It is at non-Morse points where CT shows all its power. Let us
illustrate the CT program starting with the following expression (see
Ref.~\cite{Gilm81} for a more detailed description), 
\begin{equation}
h(\vec{x},
  \vec{\lambda})=V(\vec{x}+\vec{x}_0,\vec{\lambda}+\vec{\lambda}_0)-V(\vec{x}_0,
  \vec{\lambda}_0),
\end{equation}
where $\vec{x}_0$ is a degenerated critical point (non-Morse) for the control parameter
$\vec{\lambda}_0$. We perform a Taylor expansion in the
order parameters till k-order. The problem of {\it determinacy} consists in
getting the Taylor series that can be truncated without loss of
substantial information with respect to the original function. The
issue of finding the most general family of functions with the smallest
dimension, {\it d}, which contains the original function, is known as {\it unfolding}.
The number of parameters appearing in this
unfolding is called {\it codimension} or {\it number of essential
  parameters}. This number is connected with the number of lowest
order terms in the Taylor expansion that can be canceled out. The so
called catastrophe {\it germ} is obtained when all the unfolding terms go to zero,
i.e., all possible terms in the Taylor expansion vanish,
\begin{equation}
g(x)=h(x, 0).
\end{equation}
The final concept to be introduced is {\it transversality}. One
says that the original function, $V$ is {\it k transversal} when it is
isomorphic to the canonical form of the unfolding \cite{Post78}.

Thom's {\it splitting lemma} \cite{Thom75}
guarantees that a smooth potential at non-Morse points can be
written as the sum of a quadratic form, associated to the subspace with
nonzero eigenvalues, plus a function containing the variables
associated to the zero eigenvalues of the Hessian matrix. The non-Morse part
of the latter is a canonical form called
catastrophe function. This function is composed by the catastrophe
germ, which only depends on the number of vanishing eigenvalues and on
the number of control parameters, and by a universal perturbation that
removes the degeneracy and makes the potential structurally
stable. The catastrophe germs and the related universal perturbations
were listed by Thom for potentials up to two variables and up to five parameters
\cite{Post78,Gilm81}. The
transformation into the canonical form only exists for a function
which is {\it k transversal}. If this is not the case, the function cannot
be treated with CT.

The first step in the CT program is to find out the critical
points of the energy surface ($\nabla E=0$). Among them, the most important
is the most degenerate one. This point is the fundamental root taking place at a definite value of the control
parameters which we will call critical values.
We next proceed making use of a Taylor expansion of the energy surface around the fundamental root.
A Taylor
expansion around such a point is also valid for the critical points that
{\it arise} from the fundamental root when the degeneracy is broken.
Depending on the degeneracy of the fundamental root the number of
extremes that can be analyzed simultaneously will change.

It is important to note here that if the original function is {\it k determined}, then the
k-order Taylor expansion can be transformed,  under the appropriate
non-linear change of variables, into a finite polynomial
that is valid in the neighborhood of the critical point and critical
control parameter,
\begin{equation}
\label{trans1}
x'_i=A_{i}+\delta_{ij}x_j+A_{i,jk}x_j x_k+A_{i,jkl}x_j x_k x_l+\ldots,
\end{equation}
%The number of canceled terms depends on the values of the control
%parameters.
where $x_i$ stand for the original and $x'_i$ for the transformed
variables. Finally, this polynomial can be written 
%transformed, through a
%inhomogeneous change of variables,
%\begin{equation}
%\label{trans2}
%x''_i=x'_i+A_i.
%\end{equation}
in  a canonical form
composed by the catastrophe germ plus a generic perturbation.
Once the properties of the catastrophe function have been established, the stability of the
potential is completely determined. In CT the set of non-Morse points
is known as bifurcation set while the ensemble of critical points with equal energy
are known as the Maxwell set. If the energy surface is in the
neighborhood of a Maxwell set, it possesses a first order phase transition. Conversely, if the potential is close to a bifurcation set, it has a second order phase transition.

%The bifurcation sets correspond to curves in the parameter space
%associated to degenerate critical points while the Maxwell sets
%constitute the locus of points for which the energy surface takes the
%same value in two or more critical points.

When the potential depends on several  variables it is important to
find out which are the variables involved in the phase transition, since they play the role
of order parameters. These variables are associated to the
subspace with vanishing Hessian eigenvalues, called  {\it bad} or {\it
  essential} variables, while there is another
set of variables related to the non-vanishing Hessian eigenvalues, called
{\it good} or {\it non-essential} variables. The potential
could be separated into a part depending on the {\it essential} variables and into
another part depending
on the {\it non-essential} ones by rewriting it in
terms of the eigenvectors of the Hessian matrix. As a consequence of the {\it splitting lemma} the potential is separated into a function of the {\it essential} variables and a sum of quadratic
terms associated to
the  {\it non-essential} variables \cite{Gilm81}.
Therefore, the appearance of
 critical phenomena is associated exclusively with the behavior of the
{\it essential} variables.

In  next section this program is applied to the two-fluid nuclear
IBM-2. It will be shown that the relevant elementary catastrophe for this model
is the {\it cusp catastrophe} ($A_{+3}$) which is characterized by one state variable ($z$) and
two-control parameters ($a,b$). The germ of this catastrophe is $z^4$
and the perturbation $a z + b z^2$.

%\subsection{Regions of interest}
%\label{sec-catast-regions}

\section{Application of the catastrophe theory program to IBM-2}

Even for the restricted Hamiltonian (\ref{HQ}) it is not possible
to carry out a general analysis of the whole phase diagram due to the large number of shape variables (four in
the case of the IBM-2). Moreover, it is a nontrivial task the identification of the appropriate order parameters.
In order to proceed with the analysis we will concentrate in the critical surfaces depicted in Fig.~\ref{FIG5}, which were already studied numerically in \cite{Ar04,CI04,Caprio05}.
%\bigskip
%\noindent
\subsection{Spherical-axially deformed energy surface}
%\medskip

This surface is marked with points $x-e-\overline{x}-x^*-x$ in Fig.~\ref{FIG5}. It corresponds to a situation in which $\gamma_{\pi,\nu}$ can
be assumed to be zero, because in the
spherical phase the energy does not depend on $\gamma$, and in the
axially deformed phase $\gamma=0$ (or $\gamma=\pi/3$ in the
oblate side). This assumption is not valid in the case of the line $e-x^*$ as it will be explained later on.

Following the procedure described above, we use
the fundamental root corresponding to $\beta_\pi=\beta_\nu=0$
and  $\gamma_\pi=\gamma_\nu=0$ for $\chi<0$
($\gamma_\pi=\gamma_\nu=\pi/3$ for $\chi>0$), to
construct the Hessian matrix associated to
Eq.~(\ref{Ener})
\begin{equation}
\label{hess1}
\cal{H}=
\left (
\begin{array}{cc}
\partial^2E/\partial\beta_{\pi}^2
& \partial^2E/\partial\beta_{\pi}\partial\beta_{\nu}\\
\partial^2E/\partial\beta_\nu\partial\beta_\pi & \partial^2E/\partial\beta_\nu^2
\end{array}
\right )=
\left (
\begin{array}{cc}
3\xi-2 & 2\xi-2\\
2\xi-2 & 3\xi -2
\end{array}
\right ).
\end{equation}
The two eigenvalues are $5\xi-4$ and $\xi$, and the corresponding eigenvectors
eigenvectors are,
\begin{eqnarray}
\beta_1&=& \frac{1}{2}~ (\beta_\pi+\beta_\nu),\\
\beta_2&=& \frac{1}{2}~(-\beta_\pi+\beta_\nu).
\end{eqnarray}
The eigenvalue associated to $\beta_1$ vanishes for $\xi=4/5$
while the one associated to $\beta_2$ only vanishes for the trivial
case $\xi=0$. Therefore the {\it essential} variable turns out to be $\beta_1$, while
$\beta_2$ becomes the {\it non-essential} one. If we
make an expansion of the energy in terms of $\beta_1$ and $\beta_2$ we get:
\begin{eqnarray}
\nonumber E&=& (5\xi-4)\beta_1^2 +
   \frac{4 \sqrt{2} (1-\xi) \chi}{\sqrt{7}}\beta_1^3
+\Big(8 - 9\xi  - \frac{2}{7}
      (1-\xi) \chi^2\Big) \beta_1^4\\
&+&
        \Theta(\beta_1^5) + \xi~
   \beta_2^2+\Theta(\beta_1\beta_2^2, \beta_2\beta_1^2),
\label{sphdef1}
\end{eqnarray}
where the terms $\Theta(\beta_1^5)$ and $\Theta(\beta_1\beta_2^2,
\beta_2\beta_1^2)$ can be canceled through a
nonlinear transformation (\ref{trans1}) in the {\it non-essential}
variable. Notice that the expansion (\ref{sphdef1}) has no quadratic term proportional to
$\beta_1\beta_2$ since $\beta_1$ and $\beta_2$ are eigenvectors of the
Hessian matrix. Therefore, $E$ simplifies to
\begin{equation}
 E= (5\xi-4){\beta_1}^2 +
   \frac{4 \sqrt{2} (1-\xi) \chi}{\sqrt{7}}{\beta_1}^3
   + \Big( 8 - 9\xi  - \frac{2}{7}(1-\xi)~\chi^2\Big) {\beta_1}^4+ \xi~\beta_2^2.
\label{sphdef2}
\end{equation}

In order to keep the notation simple we keep the
variable $\beta_2$, although there, it corresponds to the transformed
variable (see Eq.~(\ref{trans1})).
The most salient feature of equation (\ref{sphdef2}) is the existence
of a cubic term, which guarantees that the phase transition around $\xi=4/5$ will always be
of first order if $ \chi=\frac{\chi_\pi  +\chi_\nu}{2} \neq 0$
\cite{Gilm81}. Eq.~(\ref{sphdef2}) can be transformed into the {\it
  cusp catastrophe} for which a first order phase transition exists if the
linear term is different from zero, which is indeed the case for
$\chi\neq 0$.

Following Ref. \cite{Lope96}, the critical value of the control parameter $\xi_c$ is the solution of the equation:

\begin{equation}
r_1=-\frac{1}{2}-\frac{1}{2}\sqrt{1+\frac{r_2^2}{2}},
\end{equation}
where
\begin{equation}
r_1=\frac{35\xi_c-28}{28+4\chi^2(\xi_c-1)-21\xi_c}
\label{r1}
\end{equation}
and
\begin{equation}
r_2=\frac{8\sqrt{14}\chi(\xi_c-1)}{28+4\chi^2(\xi_c-1)-21\xi_c}.
\label{r2}
\end{equation}
Which leads to the solution:
\begin{equation}
\xi_c=\frac{28+2\chi^2}{35+2\chi^2}.
\label{xc}
\end{equation}
This expression gives the well known values $\xi_c=4/5$ for $\chi=0$
and $\xi_c=9/11$ for $\chi=\pm\sqrt{7}/2$, which are also valid for IBM-1. It is
important to note that ~(\ref{xc}) is independent on
$\chi'$ which implies that the first order IBM-1 critical line
propagates vertically generating a first order critical surface
separating spherical and axially deformed shapes in IBM-2, as was
already established in \cite{Caprio05} using different arguments.

\subsection{The $e-x^*$ line}

This line is a limiting case of the previous energy surface, defined 
by $\chi=0$ ($\chi_\pi=-\chi_\nu$). The line has some important
differences, which deserve a particular analysis. Along this line $\beta_\pi=\beta_\nu=\beta$. Moreover,
$\gamma_\pi=\pi/3-\gamma_\nu=\gamma$ cannot be zero
because one of the phases is triaxial while the other is $\gamma$
independent. These conditions define the fundamental root be
$\beta_\pi=\beta_\nu=\beta=0$, $\gamma_\pi=\gamma_\nu=\gamma=\pi/6$.

The Hessian matrix evaluated at the fundamental root can be written as,
\begin{equation}
\label{hess2}
\cal{H}=
\left (
\begin{array}{cc}
\partial^2E/\partial\beta^2
& \partial^2E/\partial\beta\partial\gamma\\
\partial^2E/\partial\gamma\partial\beta & \partial^2E/\partial\gamma^2
\end{array}
\right )=
\left (
\begin{array}{cc}
5\xi-4 & 0\\
0     & 0
\end{array}
\right ).
\end{equation}
The eigenvalue associated to the variable $\beta$ is $5\xi-4$ and it is
canceled for $\xi=4/5$. All derivatives in $\gamma$ at the fundamental root vanish.
Thus $\beta$ will be the {\it essential} variable and $\gamma $ the {\it non-essential} one.

The expansion around the fundamental root gives
\begin{equation}
\label{sphtriax1}
E=(5\xi-4) \beta^2 + (8-9\xi)\beta^4 + \Theta(\beta^6),
\end{equation}
where the odd powers vanish.
%The terms $\Theta(\beta^6)$ can be
%canceled through the nonlinear transformation (\ref{trans1}). The
%catastrophe function results,
%\begin{equation}
%\label{sphtria2}
%Cat=(1-5r) \tilde{\beta}^2 + (-1+9r)\tilde{\beta}^4,
%\end{equation}
%where $\tilde{\beta}$ is the new variable.
Since Eq.~(\ref{sphtriax1}) does not depend on $\chi$ and it has no
cubic term, the whole line $e-x^*$ will be second order
\cite{Gilm81}.

\subsection{Axially deformed-triaxial surface}

This surface is delimited by the points $e-O(6)-y-x^*-e$
($e-O(6)-\overline{y}-\overline{x}^*-e$ in the oblate side) in Fig.~\ref{FIG5}.
It represents the most complex situation because the four shape
variables ($\beta_\pi$, $\gamma_\pi$,
$\beta_\nu$, $\gamma_\nu$) have to be
treated simultaneously. With the exception for the $y$ point, the $y-O(6)$ line, and the $e-O(6)$ line, which will be treated separately, no simplification is possible.
% that will be studied later.
%On the other hand till now has been impossible to get
%analytically the position of the critical surface due to the
%complexity of the equation.
%Making use of CT and of the transtormations
%(\ref{trans1}) and (\ref{trans2}) we will stablish the order of the
%phase transition.

The fundamental root corresponds to $\gamma_\pi=\gamma_\nu=0$
($\gamma_\pi=\gamma_\nu=\pi/3$ for $\chi>0$).  The
critical values of $\beta_\pi$ and $\beta_\nu$ depend on the value of
the Hamiltonian parameters, therefore we will proceed to study the Hessian matrix for
$\gamma_\pi=\gamma_\nu=0$. The main feature of this matrix is that
%\begin{equation}
%\frac{\partial^2 E}{\partial \beta_\pi \partial \beta_\nu}=0
% \label{hes1}
%\end{equation}

\begin{equation}
\label{hes2}
\frac{\partial^2 E}{\partial \beta_\rho \partial \gamma_{\rho'}}=0,
\end{equation}
for $\gamma_\pi=\gamma_\nu=0$ ($\gamma_\pi=\gamma_\nu=\pi/3$ for
$\chi>0$) where $\rho$ and $\rho'$ stand for $\pi,
\nu$. This equation implies that
$\beta$'s are decoupled from the angular variables. As shown in
Ref.~\cite{Ar04} the behavior of $\beta$
variables when crossing the axially deformed-triaxial surface is
smooth and they are related to the non vanishing eigenvalues of the
Hessian matrix. Therefore, the $\beta$'s are {\it non-essential} variables. According to the {\it splitting lemma} the energy
surface can be expanded as a quadratic form in $\beta$'s (except in the line
$e-x^*$ already discussed) plus an expansion in the $\gamma$ variables. Note that the
coefficients will depend on the
equilibrium values of $\beta_\pi$ and $\beta_\nu$,  i.e., $\beta_\pi^0$
and $\beta_\nu^0$,
\begin{eqnarray}
\nonumber
E&=& f_{00}(\beta_{\pi}^0, \beta_{\nu}^0, \xi, \chi, \chi')+
f_{11}(\beta_{\pi}^0, \beta_{\nu}^0, \xi, \chi, \chi')\gamma_\pi
\gamma_\nu
+
f_{02}(\beta_{\pi}^0, \beta_{\nu}^0, \xi, \chi, \chi')\gamma_\nu^2
+
f_{20}(\beta_{\pi}^0, \beta_{\nu}^0, \xi, \chi, \chi')\gamma_\pi^2\\
\nonumber
&+&
f_{04}(\beta_{\pi}^0, \beta_{\nu}^0, \xi, \chi,\chi')\gamma_\nu^4 +
f_{13}(\beta_{\pi}^0, \beta_{\nu}^0, \xi, \chi,\chi')\gamma_\pi
\gamma_\nu^3 +
f_{22}(\beta_{\pi}^0, \beta_{\nu}^0, \xi, \chi, \chi')\gamma_\pi^2
\gamma_\nu^2 \\
\nonumber
&+&
f_{31}(\beta_{\pi}^0, \beta_{\nu}^0, \xi, \chi, \chi')\gamma_\pi^3
\gamma_\nu^1
+
f_{40}(\beta_{\pi}^0, \beta_{\nu}^0, \xi, \chi_\pi, \chi_\nu)\gamma_\pi^4
+ g_{20}(\beta_{\pi}^0, \beta_{\nu}^0,\xi,\chi,\chi')
(\beta_{\pi}-\beta_{\pi}^0)^2 \\
%\nonumber
&+&
g_{02}(\beta_{\pi}^0, \beta_{\nu}^0,\xi,\chi,\chi')
(\beta_{\nu}-\beta_{\nu}^0)^2
+
g_{11}(\beta_{\pi}^0,
\beta_{\nu}^0,\xi,\chi,\chi')(\beta_{\pi}-\beta_{\pi}^0)(\beta_{\nu}-\beta_{\nu}^0)
+\Theta(\gamma^5)+\Theta(\beta^3).
\label{axtriax1}
\end{eqnarray}
$\Theta(\gamma^5)$ and $\Theta(\beta^3)$ contain terms with
powers in $\gamma$ and $\beta$ equal or higher than $5$ and $3$,
respectively. The $f_{ij}$ matrix contains the coefficients
multiplying the $\gamma$'s variables, while the $g_{ij}$ matrix those multiplying the $\beta$'s variables.

If we make the expansion in terms of the eigenvalues of the Hessian matrix,
$\gamma_1, \gamma_2$ ($\beta_1, \beta_2$), the term
$\gamma_1\gamma_2$ ($\beta_1\beta_2$) in Eq.~(\ref{axtriax1}) will vanish. In what follows
 we assume that $\gamma_1$ has a vanishing eigenvalue while $\gamma_2$
has a non-vanishing one. In the case of $\beta$ both
eigenvalues are different from zero.
Next we perform a nonlinear transformation (\ref{trans1})
in $\gamma_2$, $\beta_1$ and $\beta_2$ in order to
annihilate every cross term and higher order terms. Due to the structure of
(\ref{axtriax1}), after the transformation (\ref{trans1}) the energy
surface will result in
\begin{eqnarray}
\nonumber
E&=& {f}_{00}
           (\beta_{\pi}^0, \beta_{\nu}^0, \xi, \chi, \chi')+
\tilde{f}_{20}
      (\beta_{\pi}^0, \beta_{\nu}^0, \xi, \chi, \chi')
\gamma_1^2
+
\tilde{f}_{40}(\beta_{\pi}^0, \beta_{\nu}^0, \xi, \chi,\chi')\gamma_1^4
\nonumber\\
&+&
\tilde{f}_{02}
(\beta_{\pi}^0, \beta_{\nu}^0, \xi, \chi, \chi)\gamma_2^2
+\tilde{g}_{20}(\beta_{\pi}^0,
\beta_{\nu}^0,\xi,\chi,\chi')(\beta_{1}-\beta_{1}^0)^2\nonumber\\
&+&
\tilde{g}_{02}(\beta_{\pi}^0, \beta_{\nu}^0,\xi,\chi,\chi')(\beta_{2}-\beta_{2}^0)^2
,
\label{axtriax2}
\end{eqnarray}
where that $\gamma_2$, $\beta_1$, and $\beta_2$ stand for the
transformed variables (see Eq.~\ref{trans1}).
The coefficients $\tilde{f}_{20}$ and $\tilde{f}_{02}$ are the
eivenvalues of the matrix $((f_{20},f_{11}),(f_{11},f_{02}))$, while
$\tilde{g}_{20}$ and $\tilde{g}_{02}$ are the
eivenvalues of $((g_{20},g_{11}),(g_{11},g_{02}))$.
The absence of cubic term in the later equation, identifies
this critical surface as second order
\cite{Gilm81}. Eq.~(\ref{axtriax2}) is equivalent to the {\it
  cusp catastrophe} without linear term which leads to the existence of
a unique second order phase transitions.

\subsection{The $e-O(6)$ line}
This line is a limiting case of the axially deformed-triaxial surface
analyzed in the preceding subsection. However, due to the constraints that
can be applied to the order parameters it deserves to be discussed
separately. The energy can be written in terms of a
Taylor expansion in $\beta$ for two reasons. On the one hand, we will stay in the plane with
$\chi_\pi=\chi_\nu=\chi$, i.e., $\chi'=0$. This leads to
$\beta_\pi=\beta_\nu=\beta$ and $\gamma_\pi=\gamma_\nu=\gamma$. On the other
hand, because either $\gamma=0$ (for $\chi<0$) or $\gamma=\pi/3$ (for
$\chi>0$), its influence can be absorbed in the $\beta$ variable, in
such a way that $\beta>0$ corresponds to $\gamma=0$ while $\beta<0$ to
$\gamma=\pi/3$. Therefore, the expression for the energy reduces to 
\begin{equation}
\nonumber E= (5\xi-4)\beta^2 +
   \frac{4 \sqrt{2} (1-\xi) \chi}{\sqrt{7}}\beta^3
+\Big(8 - 9\xi  - \frac{2}{7}
      (1-\xi) \chi^2\Big) \beta^4+
        \Theta(\beta^5),
\label{pro-obla}
\end{equation}
where $\Theta(\beta^5)$ can be canceled through a nonlinear
transformation in $\beta$ (\ref{trans1}).
In this case we are interested in $5\xi-4< 0$ while
$\chi$ vanishes. In this situation $\beta=0$ is a maximum while
$\beta=\pm\sqrt{\frac{4-5\xi}{16-18\xi}}$ are two degenerated
minima (note that $\beta$ corresponds to the transformed variable (see
Eq.~(\ref{trans1})).
%Note that this expression is valid in the neighbourhood of
%$\xi=4/5$.
Therefore the  $e-O(6)$ line will be first order as was
already established in \cite{Joli01}.

\section{Summary and conclusions}

In this work we introduce the key ingredients of catastrophe theory and we describe the basic program to analyze the stability and to classify the order of the phase transitions that can be developed in a
given potential energy. We have applied
the program to the case of the IBM-2 using a restricted Hamiltonian
which is of great interest in Nuclear Physics. Following this
procedure we have been able to 
determine analytically the order of the phase transitions. Our analytic results confirm previous numerical studies. In particular, we establish in an unambiguous way that the surface $x-e-\overline{x}-x^*-x$ (spherical-axially deformed surface)
is first order except for the line
$e-x^*$, which is second order. The surfaces $e-O(6)-y-x^*-e$ and
$e-O(6)-\overline{y}-x^*-e$ (axially
deformed-triaxial surfaces) are second order, except for the line
$e-O(6)$ which is first order. The relevant catastrophe for the IBM-2
phase diagram is the {\it cusp catastrophe}.

The IBM-2 example that we have treated shows that the use of catastrophe theory combined
with numerical calculations is able to determine unambiguously the
different critical potential energy surfaces present in a phase
diagram as well as the order of the phase transitions.

\section{Acknowledgments}
We thank fruitful discussions with J.\ Margalef-Roig and S.\
Miret-Art\'es.  This work was partially supported by the
Spanish Ministerio de Econom\'{\i}a y Competitividad and the European
regional development fund (FEDER) under Project Nos.
FIS2011-28738-C02-01, FIS2011-28738-C02-02, FIS2012-34479, and by Junta de
Andaluc\'{\i}a under Project Nos.\ FQM160, P11-FQM-7632 and FQM318, as well as
by Spanish Consolider-Ingenio 2010 (CPANCSD2007-00042).

%\bibliography{/users/home/jegramos/textos/biblio/refer}

\end{document}